\documentclass{svproc}
\usepackage[mathscr]{eucal}
\usepackage{amscd}
\usepackage{amssymb,latexsym}
\usepackage{graphicx}
\usepackage{amsmath}
\usepackage[utf8]{inputenc}
\usepackage{color}

\usepackage{url}

\usepackage{hyperref}

\usepackage[normalem]{ulem}

\renewcommand\l{\lambda}

\newcommand\bbR{{\mathbb R}}

\newcommand\wt{\widetilde}

\newcommand\s{\sigma}
\renewcommand\d{\partial}

\newcommand\D{\nabla}
\newcommand\e{\varepsilon}
\renewcommand\b{\beta}

\newcommand\ric{{\rm Ric}}
\renewcommand\l{\lambda}
\newcommand\g{\gamma}

\newcommand\beq{\begin{equation}}
\newcommand\eeq{\end{equation}}
\newcommand\ben{\begin{enumerate}}
\newcommand\een{\end{enumerate}}
\newcommand\bit{\begin{itemize}}
\newcommand\eit{\end{itemize}}

\newcommand{\R}{\mathbb R}

\newcommand{\pd}{\partial}

\def\undertilde#1{\mathord{\vtop{\ialign{##\crcr
   $\hfil\displaystyle{#1}\hfil$\crcr\noalign{\kern1.5pt\nointerlineskip}
   $\hfil\tilde{}\hfil$\crcr\noalign{\kern1.5pt}}}}}

\title{Remarks on the existence of \\ CMC Cauchy surfaces}

\author{Gregory J. Galloway\inst{1}  \and
Eric Ling\inst{2}}
\authorrunning{Gregory J. Galloway and Eric Ling} % abbreviated author list (for running head)
\institute{University of Miami, Coral Gables FL, USA\\
\email{galloway@math.miami.edu}
\and
Rutgers University,
New Brunswick, NJ \\
\email{eling@math.rutgers.edu}
}

\begin{document}
%\date{}
\maketitle
\vspace{.05in}

\begin{abstract}
As is well known, constant mean curvature (CMC) spacelike hypersurfaces play an important role in solving the Einstein equations, both in solving the contraints and the evolution equations.  In this paper we review the CMC existence result obtained by the authors in \cite{GalLing} and consider some new existence results motivated by a conjecture of Dilts and Holst \cite{Dilts}. We also address some issues concerning the conformal structure of cosmological spacetimes.  
\end{abstract}

\noindent

\section{Introduction}

As is well known, in order to solve the Einstein equations of general relativity, one must first obtain initial data that satisfies the Einstein constraint equations, and then use the Einstein evolution equations to evolve that data.   
%The problem of solving the constraint equations has been widely studied.  
The standard conformal method for solving the constraint equations involves solving certain equations which decouple when the mean curvature associated to the initial data is assumed to be constant.  This assumption then substantially simplifies the problem of solving the constraint equations; see e.g. \cite{Isenberg}.  There are also known advantages for solving the Einstein evolution equations if one works in CMC gauge, given CMC initial data; see e.g. \cite{Andersson}.  

One is then naturally led to consider the circumstances under which a given spacetime  admits a CMC spacelike hypersurface.  In \cite{Dilts}, Dilts and Holst address this issue for the class of what they call  {\it cosmological spacetimes}.  These are globally hyperbolic spacetimes $(M,g)$, with compact Cauchy surfaces, that satisfy the strong energy condition, $\ric(X,X) \ge 0$ for all timelike vectors  $X$.   
In this setting, they review various results that establish the existence of CMC Cauchy surfaces under various conditions (e.g. as in \cite{Gerhardt}, \cite{Bart84} \cite{Bart88}), as well as examples of spacetimes (suitably global) that fail to have CMC Cauchy surfaces (e.g.  as in \cite{Bart88}, \cite{CIP}).   Consideration of these results and examples led them  to formulate several conjectures concerning the existence (and nonexistence) of CMC Cauchy surfaces in cosmological spacetimes.  Motivated by some of their considerations, in \cite{GalLing} we obtained a new CMC existence result which relies on a certain spacetime curvature condition.  We review this result in the next section.  In addition, we present some further existence results that pertain to one of their conjectures, specifically \cite[Conjecture 3.7]{Dilts}.  These latter results also relate to a conjecture made in \cite{GalLing}, as we comment upon in the next section.  In Section 3 we address some issues related to the conformal structure of cosmological spacetimes raised in \cite{Muller}.

\medskip
\section{Some CMC existence results}

We recall some basic definitions and facts.   
%For causal theoretic notions used, but not defined below, we refer the reader to the standard references \cite{BEE, HE, ON}.  
By a spacetime we mean a smooth time oriented $(n+1)$-dimensional Lorentzian manifold $(M,g)$. 
 We further restrict to spacetimes $(M,g)$ that are globally hyperbolic.  Classically, this means (i) the `causal diamonds' $J^+(p) \cap J^-(q)$ are compact for all  $p,q \in M$ and  (ii) $M$ is strongly causal.   
Although there are slight differences in the literature, here we define a Cauchy (hyper)surface in $M$ to be an achronal set $S  \subset M$ that is met by every inextendible causal curve.  It is a classical fact that $(M,g)$ is globally hyperbolic if and only if it admits a Cauchy surface.   By considering the flow of a timelike vector field, one sees that any two Cauchy surfaces are homeomorphic, and that if $S$ is a Cauchy surface in $M$ then $M$ is homeomorphic to $\bbR \times S$.  

Let $V$ be a smooth spacelike hypersurface in a spacetime $(M,g)$.  To set conventions, the second fundamental form $K$ of $V$ is defined as:  $K(X,Y) = g(\nabla_Xu,Y)$,
where $X,Y \in T_pV$, $\D$ is the Levi-Civita connection of $M$, and $u$ is the future directed timelike unit normal vector field to $V$. Then the mean curvature $H$ is given by, $H = {\rm tr}_hK$, where $h$ is the induced metric on $V$.  By a {\it CMC Cauchy surface},  we shall always mean a smooth spacelike Cauchy surface with constant mean curvature.

We make a few comments about curvature conditions.  
We say that $(M,g)$ satisfies the strong energy condition (a.k.a.\ the timelike convergence condition)  provided
$\ric(X,X) \ge 0$ for all timelike vectors  $X$.  
Recall that if $u = e_0$ is a unit timelike vector in $(M,g)$, then $\ric(u,u)$ can be written as (minus) the sum of timelike sectional curvatures: Extend $u = e_0$ to an orthonormal frame $\{e_0, e_1, ..., e_n\}$.  Then,
\beq
\ric(u,u) = \sum_{i = 1}^n \langle R(u,e_i)e_i,u\rangle = -\sum_{i = 1}^n K(e_0 \wedge e_i).
\eeq
where $K(e_0 \wedge e_i)$ is the sectional curvature of the timelike plane spanned by $e_0$ and $e_i$.
In particular, if $(M,g)$ has everywhere nonpositive timelike sectional curvatures then $(M,g)$ satisfies the the strong energy condition.   The condition that a spacetime has nonpositive timelike sectional curvatures is the strongest condition consistent with gravity being attractive.  As shown in \cite{GalLing}, assuming natural conditions on the pressure and density, perfect fluid filled FLRW spacetimes, and sufficiently small perturbations of them, have nonpositive timelike sectional curvatures.  

\subsection{\normalsize CMC existence result from a spacetime curvature condition}

In \cite{Bart88} Bartnik constructs an interesting example of an inextendible spacetime satisfying the strong energy condition, with compact Cauchy surfaces, which does not contain any CMC Cauchy surfaces.  His example is neither future timelike geodesically complete nor past timelike geodesically complete.  This suggests considering spacetimes that are timelike geodesically complete in at least one direction.  In \cite{GalLing}, we proved the following.

\begin{theorem}\label{CMCnonpos}
Let $(M,g)$ be a spacetime with compact Cauchy surfaces. Suppose $(M,g)$ is future timelike geodesically complete and has everywhere nonpositive timelike sectional curvatures, i.e. $K \leq 0$ everywhere. Then 
$(M,g)$ contains a CMC Cauchy surface.
\end{theorem}

We make a few comments about the proof.  The proof is basically an application of the following fundamental CMC existence result of Barnik \cite{Bart88}.  

\begin{theorem}\label{BartExist}
Let $(M,g)$ be a globally hyperbolic spacetime with compact Cauchy surfaces, which satisfies the strong energy condition.  Suppose there exists a point $p \in M$ such that the past and future `light cones'  $\d J^{-}(p)$  and $\pd J^+(p)$, respectively,  are compact.
Then there exists a  CMC Cauchy surface passing through~$p$.
%regular ($C^{2,\a}$) 
\end{theorem} 

The causality condition, $\d J^{\pm}(p)$ compact, is stated in a slightly different, but equivalent,  form in \cite{Bart88}.
As they are compact achronal hypersurfaces in a globally hyperbolic spacetime, the sets $\d J^{\pm}(p)$ are necessarily Cauchy surfaces.
Theorem~\ref{CMCnonpos} is an immediate consequence of Theorem \ref{BartExist} and the following result in  \cite{GalLing}.

\begin{proposition}\label{prop for main}
Let $(M, g)$ be a spacetime with compact Cauchy surfaces and with
everywhere nonpositive timelike sectional curvatures, $K \le 0$.   If $(M, g )$ is future
timelike geodesically complete then  there exists a point $p \in M$ such that  $\d J^{\pm}(p)$  are compact.
\end{proposition}

We comment on the proof of this proposition.  Under the given assumptions, it is shown that the future causal boundary $\mathscr{C}^+$ of $(M,g)$ consists of a single point.  Without going into formal definitions,  
$\mathscr{C}^+$ consists of a single point simply means that $I^-(\g) = M$ for all future inextendible timelike curves $\g$ in $M$.  The proof of this latter statement is by contradiction, and is ultimately a consequence of Harris's Lorentzian triangle comparison theorem \cite{Harris};  see  \cite[Proposition 4]{GalLing} and  \cite[Proposition 5.11]{horo2}.   Then, by a result of Tipler \cite{Tipler}, the Bartnik condition, 
$\d J^{\pm}(p)$ compact for some $p$, is satisfied; see \cite[Proposition 3]{GalLing}.

\smallskip

In view of Theorem \ref{CMCnonpos} and the fact that the assumption of nonpositive timelike sectional curvatures implies the strong energy condition, one might be tempted, as indeed was done in \cite{GalLing}, to make the following conjecture.

\begin{conjecture}\label{conjecture}
Let $(M,g)$ be a spacetime with compact Cauchy surfaces. If $(M,g)$ is future timelike geodesically complete and satisfies the strong energy condition then $(M,g)$ contains a CMC Cauchy surface.
\end{conjecture}

We note that the conjecture holds for those spacetimes, that, in addition, admit a future complete (in the sense of completeness of integral curves) timelike conformal Killing vector field.   Indeed, as follows from results in \cite{Costa}, in this case, the future conformal boundary $\mathcal{C}^+$ consists of a single point.  Some further comments concerning $\mathcal{C}^+$ pertinent to this remark, and to the conjecture, are made in Section 3.

In the next section we obtain a result that provides some partial support for the conjecture in a natural cosmological context.

\subsection{\normalsize CMC existence result related to a conjecture of Dilts and Holst}

In \cite{Dilts}, Dilts and Holst make the point that Bartnik's nonexistence example also rules out Cauchy surfaces having mean curvature which is either strictly positive everywhere, or strictly negative curvature (but not necessarily constant).   This, together with some other considerations, led them to make the following conjecture.  

\begin{conjecture}[\cite{Dilts}, Conjecture 3.7]
Let $(M,g)$ be a spacetime with compact Cauchy surfaces, which satisfies the strong energy condition.  If $(M,g)$ has a Cauchy surface of constant \textsl{signed} mean curvature, then it contains a CMC Cauchy surface.
\end{conjecture}

Theorem \ref{maincmcexist} below provides some partial support for this conjecture.   In that theorem one assumes the existence of a compact Cauchy surface with strictly positive mean curvature, as one would expect to exist in `big bang' spacetimes. It also assumes a certain future completeness condition, as we now describe.

\smallskip
Let $(M,g)$ be a global hyperbolic Lorentzian manifold with a compact Cauchy surface $(V, h)$. By results of Bernal and S\'anchez \cite{Sanchez}, we can assume 
\beq\label{split}
M \,=\, \R \times V \:\:\:\: \text{ and } \:\:\:\: g \,=\, -\b^2 dt^2 + h_t,
\eeq
where $(V,h)$ is isometric to the $t = 0$ slice. Here $\b$ is a positive function on $M$ and $h_t$ is a Riemannian metric on the slice $V_t = \{t\} \times V$. For each $x \in V$, let $\s_x \colon [0, \infty) \to M$ denoted the $t$-line $\s_x(t) = (t,x)$. 
Throughout this section, we will make the following assumption.

\medskip
\medskip

\noindent{\bf Assumption:} The arclength of $\s_x$ is infinite for all $x \in V$.

\medskip
\medskip

This will hold, for example, if $\b$ is bounded below by a positive constant in $J^+(V)$ (i.e. $\b(x,t) \geq c > 0$ for all $(t,x) \in J^+(V)$), for then the arclength satisfies
\[
s(t) \,=\, \int_0^t \sqrt{- \langle \s_x'(r), \s_x'(r)}dr \,=\, \int_0^t\b(r,x)dr \,\geq\, \int_0^tc dr \,=\, ct 
\] 
which implies $s(t) \to \infty$ as $t \to \infty$. The future completeness of each $\s_x$ implies that $d(V, \s_x(t)) \to \infty$ as $t \to \infty$, where $d(V,p)$ denotes the Lorentzian distance from $V$ to $p \in J^+(V)$.

Consider the future directed unit timelike vector field $u = \frac{1}{\b} \pd_t$. Define the scalar $H = \text{div}_g(u)$. Let $\{u, e_1, \dotsc, e_n\}$ be an orthonormal frame at a point in some $V_t$. Then 
\[
H \,=\, \text{div}_g(u) \,=\, -\langle \nabla_u u, u \rangle + \sum_{i = 1}^n \langle \nabla_{e_i}u, e_i \rangle \,=\, \sum_{i = 1}^n \langle \nabla_{e_i}u, e_i \rangle.
\]
Therefore $H|_{V_t}$ is the mean curvature of $(V_t, h_t)$. 

By the Raychaudhuri equation for a (not necessarily geodesic) unit timelike vector field  (see e.g.  \cite[Equation 4.26]{HE}, which extends easily to higher dimensions), we have 

\begin{equation}\label{raycha equ}
\frac{dH}{d s} \,=\, -\text{Ric}(u,u) - 2\s^2 - \frac{1}{n}H^2 + \text{div}_g(\nabla_u u).
\end{equation}
Here $dH/ds$ represents the rate of change of $H$ along a $t$-line with respect to the arclength parameter $s$ of the $t$-line. $\s^2 = \s_{ij}\s^{ij} \ge 0$ where $\s_{ij}$ is the shear tensor. The vorticity tensor vanishes since $u$ is hypersurface orthogonal.

We point out that in our setting, in which \eqref{split} holds, the last term above can be expressed in terms of $\b$, as follows,
\beq\label{lapseformula}
\text{div}_g(\nabla_u u) = \frac{1}{\beta}\Delta_{t} \beta  \,,
\eeq
where $\Delta_{t}$ is the Laplacian of $(V_t, h_t)$.  We briefly indicate the proof of this.  A basic computation shows that $\nabla_u u = \text{grad}_t \ln \b$ where 
$\text{grad}_t$ is the gradient of $(V_t, h_t)$ (see e.g.\  \cite[p.\ 129]{Frankel}).  Hence, $\text{div}_g(\nabla_uu) = \text{div}_g(\text{grad}_t \ln \beta)$.
Set $X = \nabla_uu$, and let $\{u,e_1, \dotsc, e_n\}$ be an orthonormal basis at a point of $V_t$. Then,
\[
\text{div}_g(X) \,=\, \sum_{i = 1}^n \langle \nabla_{e_i}X, e_i \rangle - \langle \nabla_u X, u\rangle. 
\]
Since $X$ is tangent to $V_t$, we have $\langle \nabla_{e_i}X, e_i \rangle = \langle D_{e_i} X, e_i \rangle$ where $D$ denotes the covariant derivative for the slice $(V_t, h_t)$.  Then, letting $\text{div}_t$ denote the divergence with respect to $(V_t, h_t)$, we obtain
\begin{align*}
\text{div}_g(X)  &= \text{div}_{t}X - \langle \nabla_u X, u \rangle = \Delta_t \ln \b + |\nabla_u u |^2 \\
&= \text{div}_t \left(\frac{\text{grad}_t \b}{\b}\right) + |\text{grad}_t \ln \b|^2 =  \frac{1}{\beta}\Delta_{t} \beta \,,
\end{align*}
which establishes \eqref{lapseformula}.

\medskip
\medskip

\begin{proposition}\label{H decays prop}
Set $f = {\rm div}_g(\nabla_uu)$ and restrict to a flow line of $u$.  If $f \in L^1\big([0, \infty\big))$ and the strong energy condition holds, then $H(s) \to 0$ as $s \to \infty$. 
\end{proposition}

\medskip

\noindent
{\it Remarks.}  Note that the condition on $f$ is trivially satisfied if $u$ is a geodesic vector field. More generally,
if $\b$ has a positive lower bound on $J^+(V)$, then, from equation (\ref{lapseformula}), we see that the condition on $f$ requires, roughly speaking, that $\Delta_{t} \beta$ decays sufficiently rapidly to the future. 

\medskip

To take a slightly more physical perspective, suppose $(M,g)$ satisfies the Einstein equations, with perfect fluid source,
$$
\ric_g -\frac12 Rg = T  := (\rho + p) \nu \otimes \nu + p g \,,
$$
where  $\rho$ is the energy density, $p$ is the pressure and $\nu$ is the one form metrically dual to $u$.  Then the formula for the covariant acceleration $\D_u u$ is given by (see e.g. \cite[Proposition 12.5]{ON}), 
$$
\D_u u = - \frac{\text{grad}_t p}{ \rho + p} \,,
$$
which, after a computation similar to the one above, implies
\[
\text{div}_g(\nabla_u u) \,=\, -\frac{1}{\rho + p}\Delta_t p +\frac{1}{(\rho + p)^2}\big\langle \text{grad}_t p, \text{grad}_t(\rho + 2p)\big\rangle.
\]
In this context, the condition on $f$ requires, roughly speaking, that the spatial variation in the pressure and density decay sufficiently rapidly to the future, relative to $\rho + p$. Moreover we see that if the spatial gradient of the pressure $p$ vanishes then $u$ is a geodesic vector field and the $L^1$ condition on $f$ holds trivially.

\medskip
\medskip

\noindent
{\it Proof of Proposition \ref{H decays prop}.}
Set $q = \text{Ric}(u,u) + 2\s^2$. Integrating (\ref{raycha equ}) gives
\begin{equation}\label{integrated raycha equ}
H(s) - H(0) \,=\, -\int_0^s \left(\frac{1}{n}H^2 +q\right) + \int_0^s f.
\end{equation}
We claim that $(\frac{1}{n}H^2 + q) \in L^1\big([0, \infty)\big)$. Suppose not. Since $q$ is nonnegative (by the strong energy condition), we have $\int_0^\infty (\frac{1}{n}H^2 + q) = \infty$.  Since $\int_0^s f \leq ||f||_{L^1}$, the above equation implies 
\[
H(s) - H(0) \,\leq\, -\int_0^s \left(\frac{1}{n}H^2 +q\right) + ||f||_{L^1},
\]
and hence $H(s) \to - \infty$ as $s \to \infty$.  Therefore there exists an $s_0$ such that $H(s) \leq -1$ for all $s \geq s_0$.  Divide equation (\ref{raycha equ}) by $H^2$ and integrate from $s_0$ to $s$ to obtain
\[
\frac{1}{H(s)} \,=\, \frac{1}{H(s_0)} + \frac{1}{n}(s -s_0) + \int_{s_0}^s \frac{q}{H^2} - \int_{s_0}^s \frac{f}{H^2}.
\]
Since $\int_{s_0}^s f/H^2 \leq \int_{s_0}^s |f|/H^2 \leq \int_{s_0}^s |f| \leq ||f||_{L^1}$, we have 
\[
\frac{1}{H(s)} \,\geq\, \frac{1}{H(s_0)} + \frac{1}{n}(s -s_0) + \int_{s_0}^s \frac{q}{H^2} - ||f||_{L^1}.
\]
Therefore there exists a sufficiently large $s$ such that $1/H(s) > 0$. Since $1/H(s_0) < 0$, 
there exists an $s_1$ such that $1/H(s_1) = 0$, but this contradicts the fact that $H$ is defined for all $s$. This proves the claim.

Thus, equation (\ref{integrated raycha equ}) implies $\lim_{s \to \infty}H(s)$ exists. Since $(\frac{1}{n}H^2 + q) \in L^1\big([0, \infty)\big)$ and $q$ is nonnegative, we have $H^2 \in L^1\big([0, \infty)\big)$. Thus $\lim_{s \to \infty}H(s) = 0$. 
\qed

\medskip
\medskip

For our Theorem \ref{maincmcexist}, we will require a kind of weak localized regularity at $t = \infty$.

\begin{definition}\label{future asy reg}
\emph{
We say $(M,g)$ is \emph{future asymptotically regular} 
with respect to $V$ if for each $p \in V$ and all $\e > 0$ there is a neighborhood $U_p \subset V$ of $p$ and a time $t_p$ such that 
\[
|H(t,x) - H(t,p)| \,<\,\e
\]
for all $x \in U_p$ and $t \geq t_p$. 
}
\end{definition}

\medskip

\noindent\emph{Remark.}  Assume $\lim_{t \to \infty}H(t,p)$ exists (e.g. as in Proposition \ref{H decays prop}), and denote the limit by $H(\infty, p)$. In general, we say $H$ is \emph{continuous} at $(\infty, p)$ if for all $\e > 0$, there is an open set $(t_p, \infty) \times U_p$ such that $|H(t,x) - H(\infty, p)| < \e$ for all $(t,x) \in (t_p, \infty) \times U_p$. If $H$ is continuous at $(\infty, p)$ for all $p \in V$, then it's easy to see that $(M,g)$ will be future asymptotically regular with respect to $V$. In this sense, Definition \ref{future asy reg} is a slightly weaker condition than that of continuity at $\infty$. 

\medskip
\medskip

\begin{proposition}\label{mean curv < C prop}
Suppose $(M,g)$ is future asymptotically regular with respect to $V$.  
Set $f  = {\rm div}_g(\nabla_u u)$. If  $f \in L^1\big([0, \infty)\big)$ along each flow line of $u$ and the strong energy condition holds, then for every $C > 0$, there is a $t$-slice with mean curvature $H < C$. 
\end{proposition}

\proof
Fix $C > 0$. Let $\e = C/2$. For all $p \in V$ there is a neighborhood $U_p \subset V$ of $p$ and a time $t_p$ such that $|H(t,x) - H(t,p)| < C/2$ for all $x \in U_p$ and $t \geq t_p$. Since $V$ is compact, the open cover $\{U_p\}_{p \in V}$ has a finite subcover $\{U_{p_1}, \dotsc, U_{p_N}\}$. Let $T = \max\{t_{p_1}, \dotsc, t_{p_N}\}$.  Then for all $t \geq T$ and for all $x \in U_{p_i}$, we have $H(t,x) < H(t,p_i) + C/2$. 

By Proposition \ref{H decays prop}, we have $H\big(t(s), p\big) \to 0$ as $s \to \infty$. Since $t \to \infty$ as $s \to \infty$, we have $H(t,p) \to 0$ as $t \to \infty$.
Therefore there exists a $T$ such that $|H(t,p_i)| < C/2$ for all $t > T$ and all $i = 1, \dotsc, N$. Combining this with the result from the first paragraph, we have $H(t,x) < C/2 + C/2 = C$ for all $x \in V$ and all $t > T$. 
\qed

\smallskip
We can now prove the following CMC existence result.

\medskip

\begin{theorem}\label{maincmcexist}
Suppose $(M,g)$ is future asymptotically regular with  respect to $V$ and $V$ has positive mean curvature. Set $f = {\rm div}_g(\nabla_u u)$. If  $f \in L^1\big([0, \infty)\big)$ along each flow line of $u$ and the strong energy condition holds, then there is a Cauchy surface with constant mean curvature. 
\end{theorem}

\proof
Let $C = \inf_{x \in V} \{H(0,x)\}$. Note that $C > 0$ since $V$ is compact and $H > 0$ on $V$.  By Proposition \ref{mean curv < C prop}, there is a $t$-slice with mean curvature $H < C$.   Theorem \ref{maincmcexist} is now a consquence of the following fundamental CMC existence result, based on the presence of suitable barriers; cf. Theorem 6.1 in \cite{Gerhardt} and Theorem~4.1 in \cite{Bart84}.\qed

%\medskip
\medskip

\begin{theorem}[\cite{Gerhardt}, \cite{Bart84}]\label{cmcexist}  Let $(M,g)$ be a spacetime with compact Cauchy surfaces.   Let $S_1$ and $S_2$ be two such Cauchy surfaces, with $S_2$ in the timelike future of $S_1$, and suppose that the mean curvature $H_1$ of $S_1$ and $H_2$ of $S_2$ satisfy
\beq
H_2 < H_0 < H_1  \,,
\eeq
for some constant $H_0$.  Then there exists a Cauchy surface $S \in J^+(S_1) \cap J^-(S_2)$ with mean curvature $H_0$. 
\end{theorem} 

\medskip
\medskip

We now extend the above to a setting that permits a positive cosmological constant. The Einstein equations with a cosmological constant $\Lambda$ can be written as 
\[
\text{Ric} \,=\, 8\pi\left(T - \frac{1}{n-1}\text{tr}_g(T) g\right) + \frac{2}{n-1}\Lambda g.
\]
Setting $\l = \frac{2}{n-1}\Lambda > 0$, we see that the appropriate substitution for the strong energy condition is 
\beq\label{ric cond}
\ric(X,X) \ge -\l \,  \quad \text{for all unit timelike vectors  }  X  \,.
\eeq
We further assume that  
\beq\label{mean cond}
H^2  \ge n \l  \quad \text{on }  J^+(V).   
\eeq
This is a nontrivial assumption (unless one has $\l = 0$, as previously discussed). Part of the rationale for this assumption comes from consideration of FLRW models. In an FLRW spacetime, the first Friedmann equation gives \cite{Gibbons}
\[
H^2 \,=\, n^2\left(\frac{a'}{a}\right)^2 \,=\, \frac{n}{n-1}16 \pi \rho + n\l - \frac{n^2}{a^2}k
\]
where $\rho = T(\pd_t, \pd_t)$. Therefore, so long as $\rho \geq 0$ and $k = 0, -1$, then (\ref{mean cond}) is satisfied.

\medskip
\medskip

\begin{proposition}\label{H decays prop2}
Set $f = {\rm div}_g(\nabla_uu)$ and restrict to a flow line of $u$.  If $f \in L^1\big([0, \infty)\big)$ and \eqref{ric cond} and \eqref{mean cond} hold, then $H(s) \to  \sqrt{n \l} $ as $s \to \infty$. 
\end{proposition}

\medskip
\medskip

This can be proved in a manner very similar to Proposition \ref{H decays prop}:

\bit
\item We have $\frac{H^2}{n} + q \ge \frac{H^2}{n}  - \l \ge  0$.  One first shows that $\frac{H^2}{n} + q \in L^1$.  Otherwise $H(s) \to -\infty$.  Now argue  similarly to the proof of Proposition \ref{H decays prop} to obtain a contradiction.
\item It follows that $H(s)$ has a finite limit as $s \to \infty$.  $\frac{H^2}{n} + q \in L^1$ implies  $\frac{H^2}{n} - \l \in L^1$.
This then implies  $H(s) \to  \sqrt{n \l} $ as $s \to \infty$.   

\eit

In like manner we now obtain the following proposition and theorem.

\medskip
\medskip

\begin{proposition}\label{mean curv < C prop2}
Suppose $(M,g)$ is future asymptotically regular with respect to $V$. Set $f  = {\rm div}_g(\nabla_u u)$. If  $f \in L^1\big([0, \infty)\big)$ along each flow line of $u$, and \eqref{ric cond} and \eqref{mean cond} hold,
then for every $C >  \sqrt{n \l}$, there is a $t$-slice with mean curvature $H < C$. 
\end{proposition}

\medskip
\medskip

\begin{theorem}\label{maincmcexist2}
Suppose $(M,g)$ is future asymptotically regular with respect to $V$ and $V$ has mean curvature $> \sqrt{n \l}$. Set $f  = {\rm div}_g(\nabla_u u)$. If  $f \in L^1\big([0, \infty)\big)$ along each flow line of $u$ and \eqref{ric cond} and \eqref{mean cond} hold, then there is a Cauchy surface with constant mean curvature. 
\end{theorem}

\section{Remarks on the conformal structure of cosmological spacetimes}

Let $(M,g)$ be an $(n+1)$-dimensional spacetime with compact Cauchy surfaces.  The proof of Theorem \ref{CMCnonpos} uses the fact that if $M$ has nonpositive timelike sectional curvatures and is future timelike geodesically complete then the future causal boundary $\mathscr{C}^+$ consists of a single point.  In view of Conjecture \ref{conjecture}, one is led to ask if this remains true if the assumption of nonpositive timelike sectional curvatures is replaced by the  strong energy condition.  The paper  \cite{Muller}  attempted to answer this question in the negative. The approach was to construct, via certain conformal transformatioms, a class of spacetimes satisfying the strong energy condition, future timelike geodesic completeness, and yet would have nontrivial 
$\mathcal{C}^+$.\footnote[1]{We appreciate the constructive discussions with the author, who has subsequently withdrawn this paper.}   Here we present two propositions in order to shed some light on this situation.

Consider the following class of warped product spacetimes, sometimes referred to as generalized Robertson-Walker spacetimes,
\beq\label{GRW}
M = (0,\infty) \times V\,, \quad g = -dt^2 + f^2(t) h  \,,
\eeq
where $(V,h)$ is a compact Riemannian manifold.  It would follow from Theorem~1 in \cite{Muller} (and its proof), that there exists a function $U = U(t) = e^{u(t)}$ such that the spacetime 
$(M,\tilde g =  e^{2u}g)$ satisfies the strong energy condition and, at the same time, is future timelike geodesically complete.  We show that this does not hold in general.  

We wish to compute the Ricci curvature in this conformally rescaled metric along the geodesic $t$-lines. The computation is facilitated by making the change of variable, $d\tau = e^u dt$, i.e.,
\beq\label{tau def}
\tau(t) = \int_0^t e^{u(t')} dt'  \,,
\eeq
so that $\tilde g$ becomes,
\beq
\tilde g = -d\tau^2 + a^2 h  \, ,\quad  a = e^u f \,,
\eeq
where $u = u(t(\tau))$ and $f = f(t(\tau))$.  Then a straight forward computation gives,
%using one of the basic warp product curvature formulas as in e.g.
\begin{align}\label{ric formula}
\widetilde{\ric}(\d_t, \d_t) &=  e^{2u} \widetilde{\ric}(\d_{\tau}, \d_{\tau})  
=  e^{2u} \left[-n \left(\frac{d^2 a}{d\tau^2}\right) \slash a^2\right] \nonumber \\
&= -n \left(u'' + \frac{f'}{f} u' + \frac{f''}{f} \right)  \, ,
\end{align}
where the derivatives in \eqref{ric formula} are with respect to $t$, $' = \frac{d}{dt}$.

Let's restrict attention to the region $J^+(V_{t_0})$, where $V_{t_0}$ is a time slice $ \{t_0\} \times V$ with $t_0 > 0$.  In order for the strong energy condition to hold along the future directed normal geodesics to $V_{t_0}$, the quantity in brackets in \eqref{ric formula} would have to be nonpositive along each such geodesic.  Consider the case in which the scale factor satisfies 
\beq\label{f condition}
%f''(t) \ge 0  \quad\text{for all } t \ge 0 \, \quad \text{and} \quad 
\int_{t_0}^{\infty}  \frac{1}{f(t')}dt'  < \infty \,
\eeq
such as, for example, $f(t) = e^t$ (or $\cosh t$ as in de Sitter space) and $f(t) = t^n$, $n > 1$.  
(Note, for these particular scale factors, $\ric(\d_t,\d_t) < 0$.)

%(If, in addition, $h$ is a flat torus, $(M,g)$ is locally isometric to de Sitter space. In fact, the universal cover is isometric to the half of de Sitter space depicted in \cite[Figure (ii), p.\ 125]{HE}.)  

In order for  $\widetilde{\ric}(\d_t, \d_t)$ to be nonnegative, $u = u(t)$ must satisfy 
the differential inequality,
\beq
u'' + \frac{f'}{f} u' \le - \frac{f''}{f} \quad \text{for all } t \ge t_0 \,.
\eeq
The initial value problem
\begin{align*}
&y'' + \frac{f'}{f} y' = - \frac{f''}{f} \\
&y(t_0) = u(t_0) \, , y'(t_0) = u'(t_0)   
\end{align*}
has the unique solution
\beq
y(t) = -\ln f + C_1 \int_{t_0}^t \frac1{f(t')} dt'  +  C_2 \,
\eeq
%where $C_1 = f(0)u'(0)$ and $C_2 = u(0)$. 
where $C_1$ and $C_2$ are determined by the initial data.  By a basic comparison result (see e.g. \cite[Theorem 16, p.\ 26]{Protter}), we conclude that 
\begin{align}
u(t) &\le -\ln f + C_1 \int_{t_0}^{t} \frac1{f(t')} dt'  +  C_2  \nonumber \\
&\le -\ln f + |C_1| \int_{t_0}^{\infty} \frac1{f(t')} dt'  +  |C_2|  \,  .
\end{align}

%\quad  t \in [t_0,\infty)

By making use of this inequality in \eqref{tau def}, it is easily seen that $\tau(t)$ has a finite limit as $t \to \infty$, provided 
$f= f(t)$ satisfies \eqref{f condition}.  Hence, each timelike geodesic orthogonal to $V_{t_0}$ is future incomplete.  The situation may be summarized as follows.

\medskip

\begin{proposition}\label{prop}
Let $(M,g)$ be a spacetime of the form \eqref{GRW} where $f(t)$ satisfies \eqref{f condition}.  Then there exists no $t$-dependent conformal factor $e^{u(t)}$ such that the conformal spacetime $(M, e^{2u} g)$ is future timelike geodesically complete and satisfies the strong energy condition.
\end{proposition}

\noindent
{\it Remark.} In fact compactness of $V$ is not needed for the proof of Proposition \ref{prop}.

\medskip
The above proposition typically involves spacetimes that are future timelike geodesically complete, but that don't satisfy the strong energy condition.  The next proposition refers to future timelike geodesically incomplete spacetimes which can satisfy the strong energy condition.

\medskip
\medskip

\begin{proposition}
Let $(M,g)$ be a spacetime given by $M = (0,1) \times V$ with metric $g = -dt^2 + h$ where $(V,h)$ is a compact Riemannian manifold with constant sectional curvature.
Then there exists no $t$-dependent conformal factor $e^{u(t)}$ such that the conformal spacetime $(M, e^{2u} g)$ is future timelike geodesically complete and satisfies the strong energy condition.
\end{proposition}

\proof
Suppose such a $t$-dependent conformal factor exists. Then the conformal metric is 
\[
\wt{g} \,=\, e^{2u(t)}(-dt^2 + h) \,=\, -d\tau^2 + f^2(\tau)h,
\]
where we have introduced the coordinate change $d\tau = e^u dt$ and defined $f(\tau) = e^{u(t(\tau))}$. Hence $(M, \wt{g})$ is an FLRW spacetime. Since the strong energy condition is equivalent to everywhere nonpositive timelike sectional curvatures for FLRW spacetimes \cite[Section 3]{GalLing}, it follows that the future causal boundary of $(M, \wt{g})$ consists of a single point \cite[Proposition 4]{GalLing}. However, the causal boundary is invariant under conformal transformations, and it is readily seen that the future causal boundary of $(M,g)$ contains infinitely many points.  
\qed

\medskip
\medskip

As noted at the beginning of this section, the structure of the future causal boundary $\mathscr{C}^+$ under the assumptions of Theorem \ref{prop for main} is completely understood: $\mathscr{C}^+$ consists of a single point.  
However, the structure of $\mathscr{C}^+$ when the assumption of nonpositive timelike sectional curvatures is replaced by the strong energy condition, still appears to be an open issue.

%\newpage
%
%\bibliographystyle{amsplain}
%\bibliography{cmcnew}

%
% ---- Bibliography ----
%

\providecommand{\bysame}{\leavevmode\hbox to3em{\hrulefill}\thinspace}
\providecommand{\MR}{\relax\ifhmode\unskip\space\fi MR }
% \MRhref is called by the amsart/book/proc definition of \MR.
\providecommand{\MRhref}[2]{%
  \href{http://www.ams.org/mathscinet-getitem?mr=#1}{#2}
}
\providecommand{\href}[2]{#2}

\end{document}